\documentclass[12pt,preprint]{aastex}
\usepackage{natbib}
\slugcomment{submitted to ApJ Letters, \today}

\begin{document}

\title{Discovery of a planetary-sized object in the scattered Kuiper belt}
\author{M.E. Brown\altaffilmark{1}, C.A. Trujillo\altaffilmark{2}, D.L. Rabinowitz\altaffilmark{3}}
\altaffiltext{1}{Division of Geological and Planetary Sciences, California Institute
of Technology, Pasadena, CA 91125}
\altaffiltext{2}{Gemini Observatory, 670 North A'ohoku Place, Hilo, HI 96720}
\altaffiltext{3}{Yale Center for Astronomy and Astrophysics, Yale University, New Haven, CT 06520}
\email{mbrown@caltech.edu, trujillo@gemini.edu, david.rabinowitz@yale.edu}

\begin{abstract}
We present the discovery and initial physical and dynamical 
characterization of the object 2003 UB313. The object is 
sufficiently bright that for all reasonable values of the albedo it is
certain to be larger than Pluto. Pre-discovery observations back to 1989 are
used to obtain an orbit with extremely small errors. The object is currently
at aphelion in what appears to be a typical orbit for a scattered Kuiper belt
object except that it is inclined by about 44 degrees from the ecliptic. 
The presence of such a large object at this extreme inclination suggests
that high inclination Kuiper belt objects formed preferentially closer to the
sun.
Observations from Gemini Observatory show that
the infrared spectrum is, like that of Pluto, 
dominated by the presence of frozen methane, though visible photometry shows
that the object is almost neutral in color compared to Pluto's extremely
red color. 2003 UB313 is likely to undergo substantial seasonal change
over the large range of heliocentric distances that it travels; Pluto at its
current distance is likely to prove a useful analog for better understanding
the range of seasonal changes on this body.
\end{abstract}


\section{Introduction}
Since the discovery of the first small objects beyond Neptune
\citep{1993Natur.362..730J} astronomers have speculated about the existence
of objects larger than Pluto in the Kuiper belt. Extrapolation of the
size distribution of smaller Kuiper belt objects (KBOs) has sometimes 
been used to attempt to estimate the numbers of such larger
objects \citep[i.e.][]{2004AJ....128.1364B}, but such estimates have proven inconclusive.
One of the goals of our ongoing survey for bright KBOs
\citep{2003EM&P...92...99T} is to find the rare objects at
the bright end of the Kuiper belt magnitude distribution. Such bright
objects are invaluable as targets for detailed physical study 
\citep{2003A&A...408L..17M, 2004Natur.432..731J, 2005A&A...437.1115D,2005ApJ...627.1057T}
in addition to being potential beacons of previously unknown populations
\citep{2004ApJ...617..645B, 2004Natur.432..598K, 2004AJ....128.2564M}.

The newly discovered KBO 2003 UB313 is currently the fourth brightest
object known in the Kuiper belt (after Pluto, 2003 FY9, and 2003 EL61)
and is currently the most distant object ever 
seen in orbit around the sun. As an object notable for its brightness, distance,
and size, 2003 UB313 is certain to be the object of intensive study. We 
present here details on its discovery, preliminary observations about
its surface characteristics, and some suggestions about physical
processes operating on this object.

\section{Discovery}
2003 UB313 was discovered in data from 21 October 2003 obtained from
our ongoing survey at 
the 48-inch Samuel Oschin telescope at Palomar Observatory. At the time
of discovery the object was moving 1.42 arcseconds per hour, slower than 
the cut-off in our main survey \citep{2003EM&P...92...99T}. Our survey
obtains three images over a 3 hour period. With typical image quality of
from 2 to 3 arcseconds, slower motions are clearly detectable, but we installed
a 1.5 arcsecond per hour lower limit to our analysis to cut down the copious
false positives at the slow end. The discovery of Sedna \citep{2004ApJ...617..645B},
with a motion of 1.75 arcseconds per hour, however, suggested a need to 
efficiently search for distant objects which would be moving at lower rates.

We have now reanalyzed all survey data with a second ("slow")
detection scheme
in addition to the standard ("fast") scheme. This slow 
scheme searches for motions
between 1 and 2 arcseconds per hour between the 
first and third image of a triplet. When a potential object is found it
checks for consistency using the second image, but motion need not be 
detected between either the first and second or second and third images.
Finally, to remove the large number of false positives generated by
stationary stars, all potential detections which are within 2 arcseconds
of a catalogued USNO star are removed without examination. The slow scheme generates 10 to 20 times
more false positives than the fast scheme, leading to approximately 1200
candidates every month. These candidates are examined by eye and are generally
quickly rejected. On occasion we also make use of the Skymorph data 
base\footnote{http://skys.gsfc.nasa.gov/skymorph/skymorph.html} to 
determine that a potentially moving candidate is, in fact, a stationary star.
In the two years worth of slow data examined
to date we have found only two real objects: Sedna (previously also found in the
fast scheme) and 2003 UB313.

The extreme brightness and slow motion of 2003 UB313 made it easy to identify
it as a transient in archival data. The object was identified in multiple 
images from the Skymorph data base and eventually found in a 1989 plate from
the UK Schmidt telescope at Siding Springs Observatory. From this 16-year
orbital arc the derived barycentric orbit using the method of \citet{2000AJ....120.3323B}
gives a semi-major axis, eccentricity, and inclination of $a=67.89 \pm 0.01$, $e=0.4378 \pm 0.0001$, and $i=43.993 \pm 0.001$, respectively. 2003 UB is
currently near aphelion at $97.50\pm0.01$ AU from the sun and will not
reach perihelion at 38.2 AU until the year 2257.

Based on the semi-major axis and eccentricity, 2003 UB313 would be classified
as a typical member of the scattered Kuiper belt \citep{morbandbrown}. The 
inclination of 44 degrees is extreme for the scattered belt, however. Only
one other otherwise unremarkable
scattered object (2004 DG77) has a confirmed orbit with an 
inclination as high. 
While initial models of the scattered
Kuiper belt \citep{1997Sci...276.1670D} were incapable of populating high inclination 
regions, recent work \citep{2005CeMDA..91..109G} suggests that a combination
of gravitational scattering, mean-motion 
resonant interaction, planetary migration, and the Kozai mechanism
may be able to place objects into orbits such as these. Additional 
simulations show that objects that are initially
in the inner part of the pre-migration disk (at distances of $\sim$20 AU)
are scattered into orbits with higher inclinations than those further out
\citep{2003EM&P...92...29G}.
We expect that, on average, these inner regions will lead to the formation
of larger objects owing to both higher nebular densities and shorter 
accretion time scales. We might therefore expect to find other large
objects at high inclination in the scattered Kuiper belt. Indeed, 
the other two recently announced scattered KBOs from our survey, 
2005 FY9 and 2003 El61, both have inclinations near 30 degrees and 
approach the size of Pluto.

\section{Spectrum}
Visible photometry of 2003 UB313 was obtained on 5, 6, and 7 January
2005 using the 1.3-meter
SMARTS telescope. Data were obtained and reduced in an identical manner
to that described in \citet{rabinowitz}.
Infrared photometry was obtained on 25 and 26 
January from the Gemini
North Observatory. No evidence for any photometric
variation was seen over the short time scale
of observation.
Table 2 gives photometric and relative reflectance values
from the visible to the near infrared. No attempt is made to
correct for solar phase effects, which are of order 0.01 magnitudes
at Pluto for a 0.5 degree phase \citep{tholenbook}.
The relative brightness of 2003 UB313
is highest in the R and I filters.  
We
find an absolute magnitude of $H_r=-1.48$, which corresponds
to a diameter of $2250\rho_r^{-1/2}$~km, where $\rho_r$ is the $R$
albedo. 
Even if the surface albedo is an unreasonably high 100\% at these wavelengths
the object has a diameter approximately that of Pluto.

Medium resolution near-infrared spectra
 were obtained on the nights of 25-27 January UT 
with the Near Infrared Imager and Spectrograph \citep[NIRI,][]{2003PASP..115.1388H} 
 instrument on the Gemini North telescope. 
The J, H and K bands were measured using 3
separate grating settings and on-source times of one, one, and two hours,
respectively. Relative reflectance was computed by
dividing the spectra by a solar analog G2V star at a similar airmass
to 2003 UB313.  Each spectra was pair subtracted to remove detector
bias, then flattened and rectified.  Bad pixels and cosmic rays were
masked out in each spectrum prior to extraction.  Extracted spectra
were rebinned to a common wavelength scale with regions affected by 
bright OH
lines masked out.  Error bars were computed from the reproducibility
of spectral data in each wavelength bin. 
Though 2003 UB313 is relatively
bright, the signal-to-noise of the spectrum is only moderate owing to the
fact that at the time of discovery the object was quickly setting in
the evening sky.
Figure 1 shows the relative reflectance of 2003 UB313, with the individual
J, H, and K spectra
scaled to match the near-infrared photometry and the relative near-infrared
colors of Pluto. 
Because of uncertainties in
spectral slope across the near-infrared, we do not regard the relative scaling
between the three separate spectra to be reliable.
The near-infrared spectrum
is dominated by absorption from CH$_4$ and 
closely resembles that of Pluto. At the current signal-to-noise and systematic 
reproducibility level, no
reliable detection is made of any other species on 2003 UB313, including, 
notably, the 2.15 $\mu$m line of N$_2$, the 1.58 $\mu$m line of CO, both
detected on Pluto \citep{1993Sci...261..745O}, and the 2.01 and 2.07 $\mu$m lines 
of CO$_2$ detected on Triton \citep{1993Sci...261..742C}.
In many cases there are potentially detections of these lines, but most are 
in spectral regions
contaminated by bright sky lines or variable sky absorptions and 
none should
be believed without additional observation and confirmation. 

One major difference between the spectrum of 2003 UB313 and that of Pluto
is that the visible region of 2003 UB313 is considerably less 
red than that of Pluto. If red visible colors on icy bodies 
are interpreted as due to irradiated complex organics, the difference 
between Pluto and 2003 UB313 is surprising given the similarity of
the methane spectra of the two bodies. A more subtle difference between the
spectra is a slight shift of the positions of the methane
absorption lines (Table 2).
On Pluto methane is a minor component dissolved as a solid solution inside
of N$_2$ ice. The isolation of the methane molecules leads to a slight
but measurable energy shift in the spectrum \citep{1997Icar..127..354Q}. The four best measured 
methane lines on 2003 UB313, in contrast, appear much closer to the positions
measured in the laboratory for pure methane than they do for methane 
incorporated into N$_2$. 

\section{Discussion}

2003 UB313 is the largest known object in orbit beyond Neptune, and, like the
second largest object, Pluto, its spectrum is dominated by absorption due
to methane. Methane ices subjected to ion and UV radiation
irreversibly break down and reassemble
into more complex hydrocarbons \citep{2003EM&P...92..291M}, leading eventually to the formation
of dark red tholins \citep{1984AdSpR...4...59K}. 
The continued presence of abundant
methane on 2003 UB313 suggests the need, as has been suggested for Pluto
 \citep{spencerbook}, for an interior source to replenish the methane. The presence of 
methane on 2003 UB313 as well as Pluto 
suggests that this process is ubiquitous in the outer 
solar system and that methane is not retained on smaller objects where
escape rates are higher \citep{traftonbook}.

The red colors and large spatial albedo variations of Pluto
have been suggested to be due to distinct
regions covered by these dark red 
tholins. At Pluto's current 
heliocentric distance, dark regions absorb enough sunlight to become too
warm for methane condensation, while the bright regions serve as methane 
cold traps, thus reinforcing any albedo contrast in existence 
 \citep{2002AREPS..30..307B}. At the 97 AU distance of 2003 UB313, however, 
even dark regions will be sufficiently cold that as methane freezes
out of the atmosphere or is replenished from the subsurface it will 
cover the entire body, lowering albedo contrasts and hiding the red
tholins. This model leads to the prediction that 2003 UB313 will have
significantly less albedo variation than Pluto and that its albedo will
be as high or higher than Pluto.

The lower temperature of 2003 UB313 may also explain the difference in 
the state of the methane. Expected subsolar surface temperatures of
a 70\% albedo body at 97 AU are $\sim$30 K. At this temperature 
the vapor pressure over pure N$_2$ ice is 420 nbar, while the vapor pressure
over pure methane ice is below a pbar \citep{spencerbook}. Unlike Pluto's
present state,
methane on 2003 UB313 is currently essentially involatile 
and will not be mixed in
the atmosphere with nitrogen. 
As 2003 UB313 moved towards aphelion over the past two centuries
nitrogen and methane may have segregated, perhaps vertically. As 2003 UB313 
moves back towards perihelion a more Pluto-like mixing may occur.

The discovery of 2003 UB313 provides a new lower temperature
laboratory for the study of 
many of the processes discussed for Pluto, including atmospheric freeze
out and escape, ice chemistry, nitrogen phase transitions, and volatile
mixing and transport.
The temperature variation from perihelion of aphelion of 2003 UB313 is
even more extreme than that on Pluto. 
Higher quality infrared spectra, which should be readily obtainable for
this moderately bright object, will be a key component of future studies.

{\it Acknowledgments:}
This research is funded by the California Institute of Technology
and is also
supported by the NASA Planetary Astronomy program (MB and DR), 
and the Gemini observatory (CT). Parts of
this research are based 
on observations obtained at the Gemini Observatory, which is
operated by the Association of Universities for Research in Astronomy,
Inc., under a cooperative agreement with the NSF on behalf of the
Gemini partnership: the National Science Foundation (United States),
the Particle Physics and Astronomy Research Council (United Kingdom),
the National Research Council (Canada), CONICYT (Chile), the
Australian Research Council (Australia), CNPq (Brazil) and CONICET
(Argentina). Gemini observations included in this work were taken as part of
program GN-2004B-Q-2.


\begin{thebibliography}{30}
\expandafter\ifx\csname natexlab\endcsname\relax\def\natexlab#1{#1}\fi

\bibitem[{{Bernstein} \& {Khushalani}(2000)}]{2000AJ....120.3323B}
{Bernstein}, G. \& {Khushalani}, B. 2000, \aj, 120, 3323

\bibitem[{{Bernstein} {et~al.}(2004){Bernstein}, {Trilling}, {Allen}, {Brown},
  {Holman}, \& {Malhotra}}]{2004AJ....128.1364B}
{Bernstein}, G.~M., {Trilling}, D.~E., {Allen}, R.~L., {Brown}, M.~E.,
  {Holman}, M., \& {Malhotra}, R. 2004, \aj, 128, 1364

\bibitem[{{Brown}(2002)}]{2002AREPS..30..307B}
{Brown}, M.~E. 2002, Annual Review of Earth and Planetary Sciences, 30, 307

\bibitem[{{Brown} {et~al.}(2004){Brown}, {Trujillo}, \&
  {Rabinowitz}}]{2004ApJ...617..645B}
{Brown}, M.~E., {Trujillo}, C., \& {Rabinowitz}, D. 2004, \apj, 617, 645

\bibitem[{{Cruikshank} {et~al.}(1993){Cruikshank}, {Roush}, {Owen}, {Geballe},
  {de Bergh}, {Schmitt}, {Brown}, \& {Bartholomew}}]{1993Sci...261..742C}
{Cruikshank}, D.~P., {Roush}, T.~L., {Owen}, T.~C., {Geballe}, T.~R., {de
  Bergh}, C., {Schmitt}, B., {Brown}, R.~H., \& {Bartholomew}, M.~J. 1993,
  Science, 261, 742

\bibitem[{{de Bergh} {et~al.}(2005){de Bergh}, {Delsanti}, {Tozzi}, {Dotto},
  {Doressoundiram}, \& {Barucci}}]{2005A&A...437.1115D}
{de Bergh}, C., {Delsanti}, A., {Tozzi}, G.~P., {Dotto}, E., {Doressoundiram},
  A., \& {Barucci}, M.~A. 2005, \aap, 437, 1115

\bibitem[{{Dout{\' e}} {et~al.}(1999){Dout{\' e}}, {Schmitt}, {Quirico},
  {Owen}, {Cruikshank}, {de Bergh}, {Geballe}, \&
  {Roush}}]{1999Icar..142..421D}
{Dout{\' e}}, S., {Schmitt}, B., {Quirico}, E., {Owen}, T.~C., {Cruikshank},
  D.~P., {de Bergh}, C., {Geballe}, T.~R., \& {Roush}, T.~L. 1999, Icarus, 142,
  421

\bibitem[{{Duncan} \& {Levison}(1997)}]{1997Sci...276.1670D}
{Duncan}, M.~J. \& {Levison}, H.~F. 1997, Science, 276, 1670

\bibitem[{{Gomes}(2003)}]{2003EM&P...92...29G}
{Gomes}, R. 2003, Earth Moon and Planets, 92, 29

\bibitem[{{Gomes} {et~al.}(2005){Gomes}, {Gallardo}, {Fern{\' a}ndez}, \&
  {Brunini}}]{2005CeMDA..91..109G}
{Gomes}, R.~S., {Gallardo}, T., {Fern{\' a}ndez}, J.~A., \& {Brunini}, A. 2005,
  Celestial Mechanics and Dynamical Astronomy, 91, 109

\bibitem[{{Grundy} \& {Fink}(1996)}]{1996Icar..124..329G}
{Grundy}, W.~M. \& {Fink}, U. 1996, Icarus, 124, 329

\bibitem[{{Hodapp} {et~al.}(2003){Hodapp}, {Jensen}, {Irwin}, {Yamada},
  {Chung}, {Fletcher}, {Robertson}, {Hora}, {Simons}, {Mays}, {Nolan}, {Bec},
  {Merrill}, \& {Fowler}}]{2003PASP..115.1388H}
{Hodapp}, K.~W., {Jensen}, J.~B., {Irwin}, E.~M., {Yamada}, H., {Chung}, R.,
  {Fletcher}, K., {Robertson}, L., {Hora}, J.~L., {Simons}, D.~A., {Mays}, W.,
  {Nolan}, R., {Bec}, M., {Merrill}, M., \& {Fowler}, A.~M. 2003, \pasp, 115,
  1388

\bibitem[{{Jewitt} \& {Luu}(1993)}]{1993Natur.362..730J}
{Jewitt}, D. \& {Luu}, J. 1993, \nat, 362, 730

\bibitem[{{Jewitt} \& {Luu}(2004)}]{2004Natur.432..731J}
{Jewitt}, D.~C. \& {Luu}, J. 2004, \nat, 432, 731

\bibitem[{{Kenyon} \& {Bromley}(2004)}]{2004Natur.432..598K}
{Kenyon}, S.~J. \& {Bromley}, B.~C. 2004, \nat, 432, 598

\bibitem[{{Khare} {et~al.}(1984){Khare}, {Sagan}, {Thompson}, {Arakawa},
  {Suits}, {Callcott}, {Williams}, {Shrader}, {Ogino}, {Willingham}, \&
  {Nagy}}]{1984AdSpR...4...59K}
{Khare}, B.~N., {Sagan}, C., {Thompson}, W.~R., {Arakawa}, E.~T., {Suits}, F.,
  {Callcott}, T.~A., {Williams}, M.~W., {Shrader}, S., {Ogino}, H.,
  {Willingham}, T.~O., \& {Nagy}, B. 1984, Advances in Space Research, 4, 59

\bibitem[{{Marchi} {et~al.}(2003){Marchi}, {Lazzarin}, {Magrin}, \&
  {Barbieri}}]{2003A&A...408L..17M}
{Marchi}, S., {Lazzarin}, M., {Magrin}, S., \& {Barbieri}, C. 2003, \aap, 408,
  L17

\bibitem[{{Moore} {et~al.}(2003){Moore}, {Hudson}, \&
  {Ferrante}}]{2003EM&P...92..291M}
{Moore}, M.~H., {Hudson}, R.~L., \& {Ferrante}, R.~F. 2003, Earth Moon and
  Planets, 92, 291

\bibitem[{{Morbidelli} \& {Brown}(2005)}]{morbandbrown}
{Morbidelli}, A. \& {Brown}, M.~E. 2005, {Comets II, ed. M.C. Festou, H.U.
  Keller, H.A. Weaver (U. Arizona Press: 2005)}

\bibitem[{{Morbidelli} \& {Levison}(2004)}]{2004AJ....128.2564M}
{Morbidelli}, A. \& {Levison}, H.~F. 2004, \aj, 128, 2564

\bibitem[{{Owen} {et~al.}(1993){Owen}, {Roush}, {Cruikshank}, {Elliot},
  {Young}, {de Bergh}, {Schmitt}, {Geballe}, {Brown}, \&
  {Bartholomew}}]{1993Sci...261..745O}
{Owen}, T.~C., {Roush}, T.~L., {Cruikshank}, D.~P., {Elliot}, J.~L., {Young},
  L.~A., {de Bergh}, C., {Schmitt}, B., {Geballe}, T.~R., {Brown}, R.~H., \&
  {Bartholomew}, M.~J. 1993, Science, 261, 745

\bibitem[{{Quirico} \& {Schmitt}(1997)}]{1997Icar..127..354Q}
{Quirico}, E. \& {Schmitt}, B. 1997, Icarus, 127, 354

\bibitem[{{Rabinowitz} {et~al.}(2005){Rabinowitz}, {Tourtellotte}, {Brown}, \&
  {Trujillo}}]{rabinowitz}
{Rabinowitz}, D., {Tourtellotte}, S., {Brown}, M., \& {Trujillo}, C. 2005, DPS
  abstract 56.12

\bibitem[{{Rudy} {et~al.}(2003){Rudy}, {Venturini}, {Lynch}, {Mazuk},
  {Puetter}, \& {Brad Perry}}]{2003PASP..115..484R}
{Rudy}, R.~J., {Venturini}, C.~C., {Lynch}, D.~K., {Mazuk}, S., {Puetter},
  R.~C., \& {Brad Perry}, R. 2003, \pasp, 115, 484

\bibitem[{{Spencer} {et~al.}(1997){Spencer}, {Stansberry}, {Trafton}, {Young},
  {Binzel}, \& {Croft}}]{spencerbook}
{Spencer}, J.~R., {Stansberry}, J.~A., {Trafton}, L.~M., {Young}, E.~F.,
  {Binzel}, R.~P., \& {Croft}, S.~K. 1997, {Pluto and Charon, ed. S.A. Stern,
  D.J Tholen (U. Arizona Press:1997)}

\bibitem[{{Tholen} \& {Buie}(1997)}]{tholenbook}
{Tholen}, D.~J. \& {Buie}, M.~W. 1997, {Pluto and Charon, ed. S.A. Stern, D.J
  Tholen (U. Arizona Press:1997)}

\bibitem[{{Trafton} {et~al.}(1997){Trafton}, {Hunten}, {Zahnle}, \& {McNutt,
  Jr.}}]{traftonbook}
{Trafton}, L.~M., {Hunten}, D.~M., {Zahnle}, K.~J., \& {McNutt, Jr.}, R.~L.
  1997, {Pluto and Charon, ed. S.A. Stern, D.J Tholen (U. Arizona Press:1997)}

\bibitem[{{Trafton} \& {Stern}(1996)}]{1996AJ....112.1212T}
{Trafton}, L.~M. \& {Stern}, S.~A. 1996, \aj, 112, 1212

\bibitem[{{Trujillo} \& {Brown}(2003)}]{2003EM&P...92...99T}
{Trujillo}, C.~A. \& {Brown}, M.~E. 2003, Earth Moon and Planets, 92, 99

\bibitem[{{Trujillo} {et~al.}(2005){Trujillo}, {Brown}, {Rabinowitz}, \&
  {Geballe}}]{2005ApJ...627.1057T}
{Trujillo}, C.~A., {Brown}, M.~E., {Rabinowitz}, D.~L., \& {Geballe}, T.~R.
  2005, \apj, 627, 1057

\end{thebibliography}

\eject

\begin{figure}
\caption{Relative reflectance of 2003 UB313 (solid points with error bars) and absolute reflectance of Pluto (red line). The large points are the reflectance
derived from BVRIJHK photometry.
Every reliably identifiable feature in the 1 - 2.5 $\mu$m region of the spectrum
of 2003 UB313 is due to absorption by solid methane. The absolute geometric
albedo of 2003 UB313 is not yet known. The relative reflectance is scaled
to match that of Pluto in the $I$ band for comparison.
The Pluto spectrum is a compilation from \citet{1996AJ....112.1212T}, \citet{1996Icar..124..329G}, \citet{2003PASP..115..484R}, and \citet{1999Icar..142..421D}.
}
\end{figure} 
\begin{figure}
\plotone{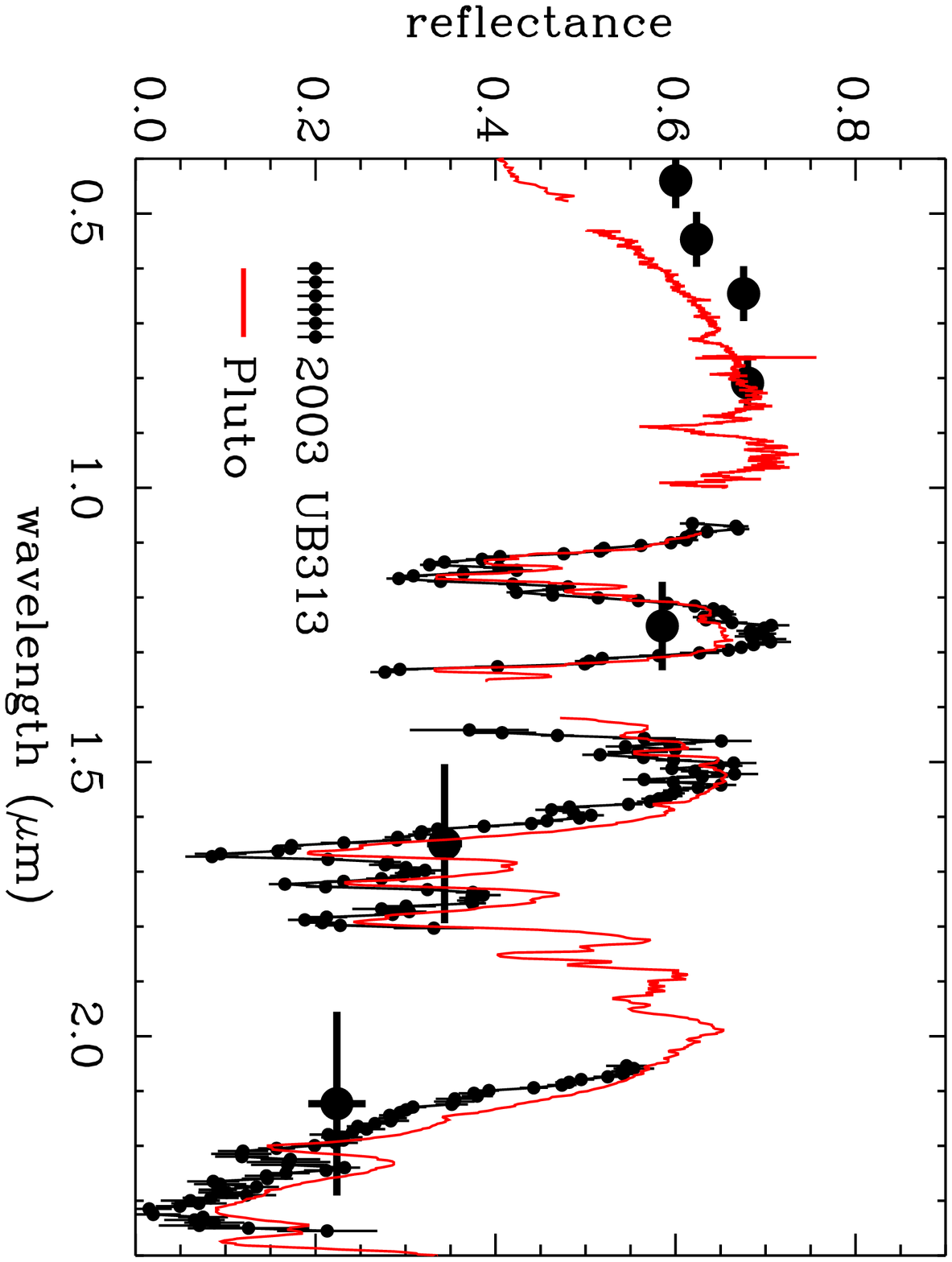}
\end{figure}
\eject
\begin{table}
\begin{center}
\caption{Photometry of 2003 UB313}
\begin{tabular}{cccc}
\tableline\tableline
filter & magnitude & relative reflectance \\
\tableline
B &$19.54\pm .01$&   0.88  \\
V &$18.83 \pm.02$&  0.92 \\
R &$18.38 \pm.02$&  1.00 \\
I &$18.05 \pm.02$&  1.00 \\
J &$17.82 \pm.02$&  0.86 \\
H &$18.11 \pm.03$&  0.51\\
K &$18.51 \pm.05$&  0.32 \\
\tableline
\end{tabular}
\end{center}
\end{table}

\begin{table}
\begin{center}
\caption{Positions of methane lines}
\begin{tabular}{cccc}

\tableline
\tableline
line identification & 2003 UB31$^a$ &pure methane$^b$ &	methane in nitrogen$^b$ \\
 & ($\mu$m) & ($\mu$m) & ($\mu$m) \\
\tableline
$\nu_2+2\nu_3+\nu_4$&	1.138&	1.139&	1.136\\
$2\nu_3+\nu_4 \hfil \hfil$	&	1.165&	1.165&	1.161\\
2$\nu_3 \hfil \hfil \hfil$	&	1.670&	1.670&	1.666\\
$\nu_2+\nu_3+\nu_4$&	1.723&	1.725&	1.720\\
\tableline
\end{tabular}
\tablenotetext{a}{Wavelength uncertainties are approximately $\pm$0.002$\mu$m}
\tablenotetext{b}{Laboratory data from \citet{1997Icar..127..354Q}}
\end{center}
\end{table}

\end{document}